\documentclass[lettersize,journal]{IEEEtran}
\usepackage{amsmath,amsfonts}
\usepackage{algorithmic}
\usepackage{algorithm}
\usepackage{array}
\usepackage[caption=false,font=normalsize,labelfont=sf,textfont=sf]{subfig}
\usepackage{textcomp}
\usepackage{stfloats}
\usepackage{url}
\usepackage{verbatim}
\usepackage{graphicx}
\usepackage{cite}
\usepackage{xcolor}
 \usepackage{multirow}

\begin{document}

\title{An Ideal Random Number Generator Based on Quantum Fluctuations and Rotating Wheel for Secure Image Encryption }

\author{Subhadip Rana, Sanku Paul, and Mrinal Kanti Mandal
\thanks{Manuscript received Month DD, YYYY; revised Month DD, YYYY; accepted Month DD, YYYY. Date of publication Month DD, YYYY; date of current version Month DD, YYYY.}
\thanks{S. Rana is with the Department of Physics, National Institute of Technology Durgapur, West Bengal, India (e-mail: sr.22ph1101@phd.nitdgp.ac.in).}
\thanks{S. Paul is with the Department of Physics and Complex Systems, S.N. Bose National Centre for Basic Sciences, Kolkata 700106, India (e-mail: sankup005@gmail.com)
}
\thanks{M. K. Mandal is with the Department of Physics, National Institute of Technology Durgapur, West Bengal, India (e-mail: mkmandal.phy@nitdgp.ac.in).}
}

\maketitle

\begin{abstract}In the era of digitization secure transmission of digital images has become essential in real world applications. Image encryption is an effective technique for protecting image data from unauthorized access. The security of encrypted data strongly depends on the quality of the random numbers used as the encryption key. In this paper, we proposed a hybrid random number generator based on quantum fluctuations and an algorithmically inspired rotating wheel. The wheel contains integer values from 0 to 255 that are shuffled using quantum fluctuations generated by time-evolving the quantum kicked rotor model. There are four pre-define tapping positions in the rotating wheel to collect the number sequences. The wheel rotation speed is dynamically varied after each set of tapping to enhance unpredictability. The entropy of the number sequence obtained from the rotating wheel attains the ideal value of 8 (in an 8 bit representation). Further, the generated  number sequences exhibit a flat histogram and nearly zero correlation, indicating strong randomness. The generated sequences are applied to the image encryption and analyzed cryptographically. Experimental results demonstrate a near ideal entropy of 7.997, an NPCR of 99.60\%, low correlation in all directions, and low PSNR for encrypted images. These results confirm that the proposed random number generator achieves efficient and high-security performance, making it suitable for the security of consumer applications such as mobile healthcare imaging, biometric authentication, QR-based and multimedia communication on smart devices.
\end{abstract}


\section{Introduction}

In the era of digitization, digital images are widely used in both daily life and specialized applications. These include social media sharing, online banking, tele-medicine, medical diagnostics, military communication, remote sensing, surveillance systems, and cloud-based multimedia services. The rapid increase in image transmission over public networks has raised serious concerns about the privacy and security of images. Digital images often contain sensitive information, so sending them safely is a very important concern in real life. Image encryption\cite{bib1} is one of the most effective techniques for protecting digital images from unauthorized access. Digital images exhibit high redundancy, strong correlation among adjacent pixels, and large data volumes. These limitation explain why conventional text\cite{bib2,bib3,bib4} oriented cryptographic algorithms are less efficient for image security. Random numbers\cite{bib5} play a fundamental role in modern cryptographic systems. The security of an encryption algorithm strongly depends on the quality of randomness\cite{bib6} used in key generation, permutation, and diffusion processes. Random number generators are generally classified into pseudo-random number generators (PRNGs)\cite{bib7,bib8} and true random number generators (TRNGs)\cite{bib9,bib10}. The PRNGs are based on deterministic mathematical models, generated via numerical simulation whereas TRNGs exploit unpredictable physical phenomena. Although PRNGs are widely used due to their simplicity and low computational cost. Their deterministic nature makes them vulnerable to various cryptanalytic, statistical, and side-channel attacks.
Chaos-based\cite{bib11,bib12,bib13} random number generators have been extensively used for image encryption applications due to their sensitivity to initial conditions and complex dynamical behavior. However, chaotic random number generators implemented using floating-point operations may exhibit CPU-dependent behavior, where variations in floating-point precision and arithmetic units introduce inconsistencies in randomness and security performance. These limits\cite{bib14} indicate the needs for efficient methodologies to generate random numbers that use only integer operations. Such mechanism should be platform independent and improve robustness against cryptanalysis. The quantum fluctuations\cite{bib15} provide an inherent source of physical randomness and have been recognized as a promising candidate for random number generation. However, purely hardware-based quantum random number generators often suffer from high corelation, limited availability, and sensitivity to environmental noise. Inspired by these challenges, this work studies a hybrid method that mixes quantum fluctuation based randomness with a rotating wheel mechanism. This approach improves randomness while keeping the computation fast. The proposed method uses a wheel with integer values from 0 to 255. These values are arranged based on quantum fluctuations. Random numbers are taken from multiple sampling points in each iteration. The wheel is rotated at changing speeds to make the numbers more unpredictable. The generated concatenated random numbers are then used for image encryption. Statistical and security tests are performed to check the quality of randomness and the strength of the encryption algorithm.
The main contributions of this work are summarized as follows:
\begin{itemize}
    \item A hybrid random number generation framework based on a rotating wheel structure, where the wheel elements (0-255) are ordered using quantum fluctuation  
    driven randomness.
    \item A key-dependent multi-tap sampling mechanism in which multiple random values are extracted simultaneously from fixed wheel positions derived from a 32 secret key.
    \item A dynamic wheel rotation strategy with incrementally varying speed enhanced temporal randomness without modifying the wheel contents.
    \item Application of the generated random sequences to digital image encryption and cryptanalysis.
    \end{itemize}
\section{Quantum fluctuation}
\begin{figure}[h]
\centering
\includegraphics[width=.9\linewidth,height=5cm]{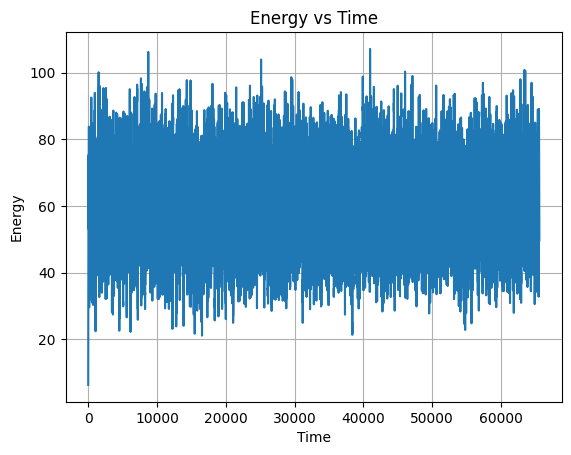}
\caption{{Energy vs time graph of the  kick rotor system with kick strength $K=5$.}}
\label{Fig-evt}
\end{figure}

In this section, we discuss the method to obtain quantum fluctuations require to order the wheel elements in the rotating wheel structure and shuffle image. To this end, we consider a quantum system composed of a particle rotating on a ring and receiving time-periodic kicks, namely the kicked rotor\cite{bib135, bib134}. It is a paradigmatic model in quantum chaos. This system has been used to study various phenomena such as transport phenomenon\cite{bib123,bib124,bib125} quantum-classical correspondence\cite{bib126,bib127}, quantum control\cite{bib128,bib129}, localization\cite{bib130} and decoherence\cite{bib131,bib132,bib133}effects. The Hamiltonian of the system is given by
\begin{equation}
    \hat{H} = \frac{\hat{p}^2}{2} + K\cos \hat{x} \sum_n \delta(t-nT)
    \label{eq_Ham}
\end{equation}
where $\hat{x}(\hat{p})$ represents the position (momentum) operator(s), $K$ is the kick strength applied at periodic intervals of time with period $T=1$ using the $\delta(t-nT)$ function. The dynamics of the systems is defined on a cylinder with $x\in [-\pi,\pi]$ and $p\in(-\infty,\infty)$. Due to the time-periodic kicking, the energy is not conserved. In principle, the system can display unbounded energy growth. Interestingly, for $K\gg 1$, the system is known to display a well-known dynamical localization phenomenon, \emph{i.e.}, energy growth is suppressed. The dynamical localization is the result of interference of different scattering momentum paths. To obtain the time evolution, start with an initial state $|\psi(0)\rangle = |p=0\rangle$. The dynamics is generated using the time-evolution operator $\hat{U}=e^{-i\hat{p}^2/2} e^{-iK\cos \hat{x}}$. The time-evolved state is obtained as $|\psi(t)\rangle = \hat{U}^t |\psi(0)\rangle$. Having $|\psi(t)\rangle$, the energy, $\langle E \rangle$, of the system at any time $t$ is evaluated as
\begin{equation}
    \langle E \rangle = \frac{1}{2}\left[\langle \psi(t)| \hat{p}^2 | \psi(t)\rangle -\langle \psi(t)| \hat{p} | \psi(t)\rangle^2\right]\,.
    \label{eq_ene}
\end{equation}
It is necessary to point out that eq. \ref{eq_ene} also describes fluctuation in momentum space. These fluctuations have quantum origin. Figure.\ref{Fig-evt} illustrates the energy versus time graph, confirming the randomness of the generated energy sequence. The linear correlation coefficient of 0.0021, together with the NIST statistical randomness test reported in Table\ref{tab:NIST}, further exhibits the strong randomness characteristics of the generated sequence.
 \\
\textbf{Index Generation:}  
Before encryption, the quantum kicked rotor system is iteratively evolved to generate an energy sequence. 
From the energy sequence, we construct a one-dimensional array
\( P = \{P(i)\}_{i=1}^{256} \), where each element takes integer values
in the range \([0,255]\). The elements of \( P \) are defined as
\begin{equation}
P(i) = \left\lfloor E_i \times 10^{4} \right\rfloor \bmod 256,
\end{equation}
where \(E_i\) denotes the \(i\)-th energy value.
In an analogous manner, two matrices \(m\) and \(n\) are generated for
image-level shuffling, with elements satisfying
\(m(i,j) \in [0, M-1]\) and \(n(i,j) \in [0, N-1]\), respectively.
Here, \(M \times N\) denotes the dimensions of the test image.
\section{Rotating Wheel Based Random Number Generation}
In this section, we describe the proposed rotating-wheel based random number generation method. The design is inspired by a mechanical rotation process. It combines key-dependent sampling with changing rotation speeds to improve randomness.
\subsection{Wheel Initialization}
The rotating wheel consisting of integer values ranging from 0 to 255. To introduce physical randomness into the system, the wheel elements are permuted using index pairs generated from quantum fluctuation based random numbers.
\noindent Let $P(i)$ denote index  of size $1\times 256$, derived from quantum fluctuation data and normalized to the range $[0,255]$. For each position $(i)$ on the wheel, the element at that location is swapped with the element at position ($P(i)$). It is illustrated in equation given below
\begin{equation}
w(i) \leftrightarrow w(P(i)).
\end{equation}
\noindent This swapping operation is performed sequentially for all positions on the wheel. As a result we get a fully permuted wheel configuration.
\begin{figure}[h]
\centering
\includegraphics[width=.65\linewidth,height=5cm]{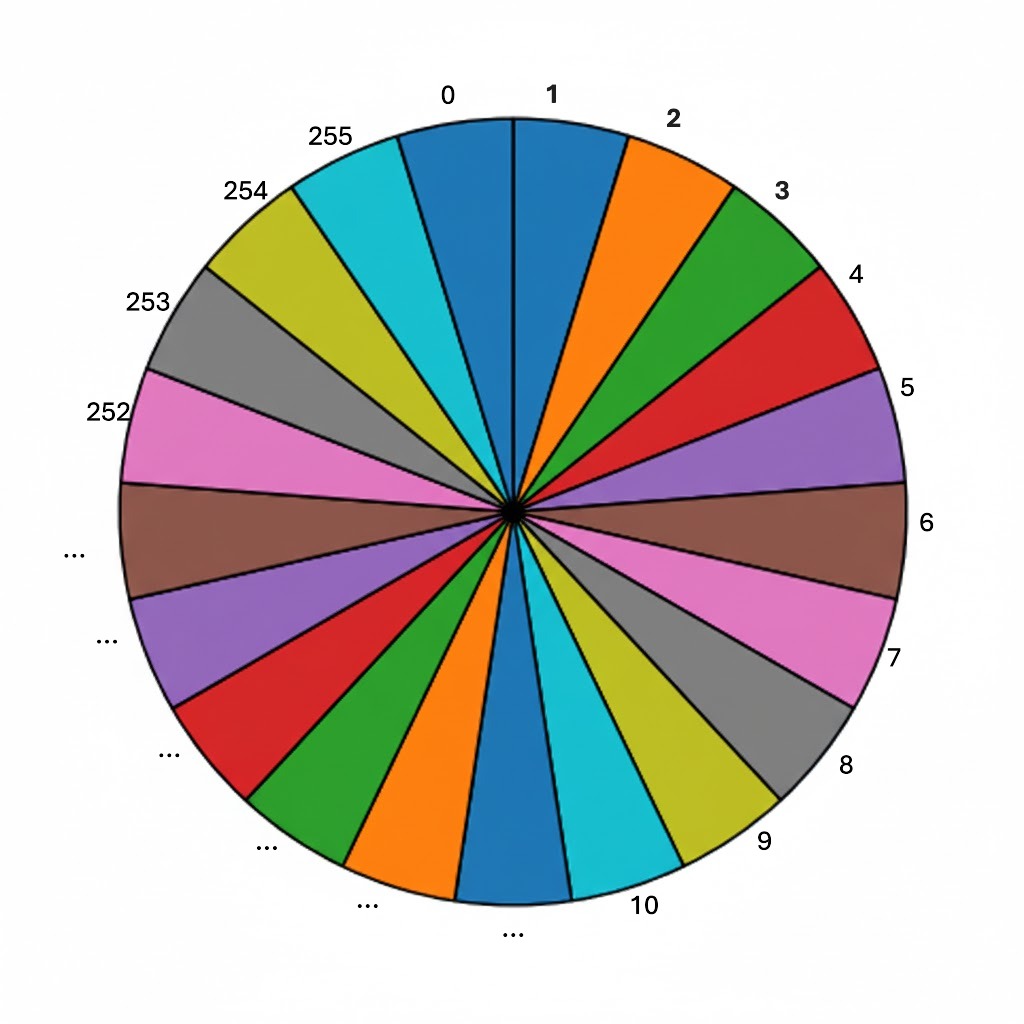}
\caption{{Wheel image replica before shuffling.}}
\label{Fig-wh}
\end{figure}

\subsection{Multi-Tap Sampling Strategy}
A sampling strategy is employed in the proposed method. At each iteration, four different positions on the rotating wheel are sampled at the same time. Each sampling operation reads the wheel value at the selected tap position, producing four random numbers per iteration. This strategy allows fast generation of a large amount of random data and reduces correlation among the generated values.
A 32-byte secret key is used to determine the sampling positions on the wheel. First, the key characters are converted into their ASCII values and accordingly this four initial tap positions are calculated. These tap positions are remain fixed throughout the generation process and act as secret parameters that control the generation of random numbers,
\begin{equation}
\mathbf{k} = \{k_1, k_2, \dots, k_{32}\},
\end{equation}
where $k_i \in \mathbb{Z}$ denotes the ASCII value of the $i$-th key character.
The four tap positions are derived from different segments of the key using summation and modular product operations. The tap positions are computed as follows.
\begin{equation}
x_0 = \left( \sum_{i=1}^{6} k_i + \prod_{i=7}^{12} k_i \bmod 9 \right) \bmod 256,
\end{equation}
\begin{equation}
y_0 = \left( \sum_{i=6}^{11} k_i + \prod_{i=12}^{17} k_i \bmod 19 \right) \bmod 256,
\end{equation}
\begin{equation}
z_0 = \left( \sum_{i=11}^{16} k_i + \prod_{i=17}^{22} k_i \bmod 83 \right) \bmod 256,
\end{equation}
\begin{equation}
w_0 = \left( \sum_{i=16}^{21} k_i + \prod_{i=22}^{27} k_i \bmod 57 \right) \bmod 256.
\end{equation}
\begin{figure}[h]
\centering
\includegraphics[width=.65\linewidth,height=5cm]{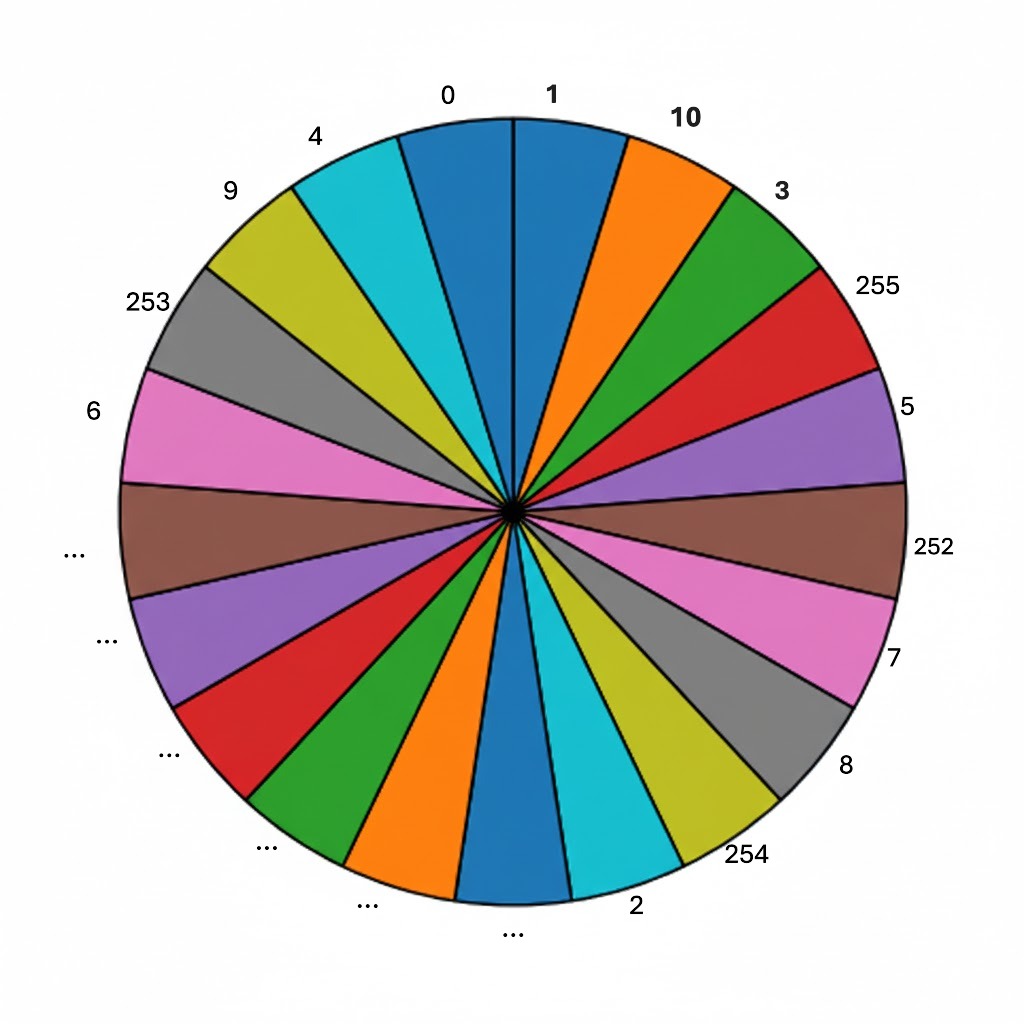}
\caption{{Shuffled Wheel image with tapping position.}}
\label{Fig-wht}
\end{figure}
\subsection{Dynamic Wheel Rotation} Initially the wheel is at rest and after collection of data from the tapping points, its speed increases with the parameter value  $\delta$ in counter clock wise direction.\[
\delta =int\frac{( x_0 \oplus y_0 )} {|x_0 - y_0|}
\]
This incremental speed variation introduces temporal randomness by ensuring that the relative alignment between the wheel contents and the fixed tap positions continuously changes over successive iterations. As a result, the generator enhances unpredictability without modifying the wheel contents.
\subsection{Random Sequence Generation} All the number sequence is obtain from tapping positions are concatenated in a 2D arrays to match the target image dimensions. 
Finally, the sequence is shuffled using the permutation matrices $m(i,j)$ and $n(i,j)$ generated from quantum fluctuations.
\begin{equation}
R(i,j) = R\big(m(i,j),\, n(i,j)\big),
\label{eq:key_permutation}
\end{equation}

\section{Encryption Algorithm}
\begin{figure}[h]
\centering
\includegraphics[width=.6\linewidth]{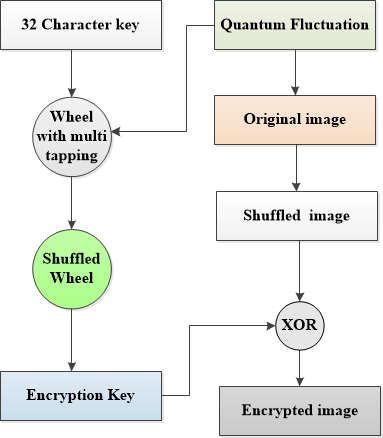}
\caption{{Block diagram of the propose algorithm.}}
\label{Fig-bd}
\end{figure}

The details of the encryption algorithm are discussed step-by-step below and the corresponding flow chart is shown in the Fig~\ref{Fig-bd}.\\
\textbf{Step-1:}  
A grayscale image $I$ of size $M \times N$ is taken as the plaintext input. The image pixels are permuted according to
\begin{equation}
I(i,j) = I\big(m(i,j),\, n(i,j)\big),
\label{eq:Image_permutation}
\end{equation}
The shuffling of plaintext image destroys  correlations among neighboring pixels and improving resistance to cropping, noise and chosen-plaintext attacks.\\
\textbf{Step-2:}  
A rotating wheel containing integer values from 0 to 255 is initialized. The wheel elements are   permuted using the index matrices $P(i)$ :
\[
w(i) \leftrightarrow w\big(P(i)\big).
\]
This step introduces key-dependent randomness into the wheel configuration using quantum-derived indices.\\
\textbf{Step-3:}  
The keystream is generated using a key-dependent multi-tap sampling from permuted wheel. \\
\textbf{Step-4:}  
The generated keystream is reshaped into a matrix of size $M \times N$ and further permuted using the same image-level index matrices as illustrated in Eq:\ref{eq:key_permutation} which strengthens confusion by preventing structural alignment between the keystream and the permuted plaintext.\\
\textbf{Step-5:}  
Finally, the permuted plaintext image is diffused with the shuffled keystream using a bitwise XOR operation:
\[
C(i,j) = I(i,j) \oplus R(i,j),
\]
producing the encrypted image $C$.\\ 
\textbf{Step-6:} This encrypted image can be communicated in public channel and can be store in cloud based system.\\
\textbf{Decryption}
Decryption is performed by applying the inverse operations in reverse order using the same secret key and index matrices. The 32-character secret key is securely transmitted through a confidential channel.

\section{Results and Discussion}\label{Rd}
\begin{figure}
\centering
\includegraphics[width=.99\linewidth]{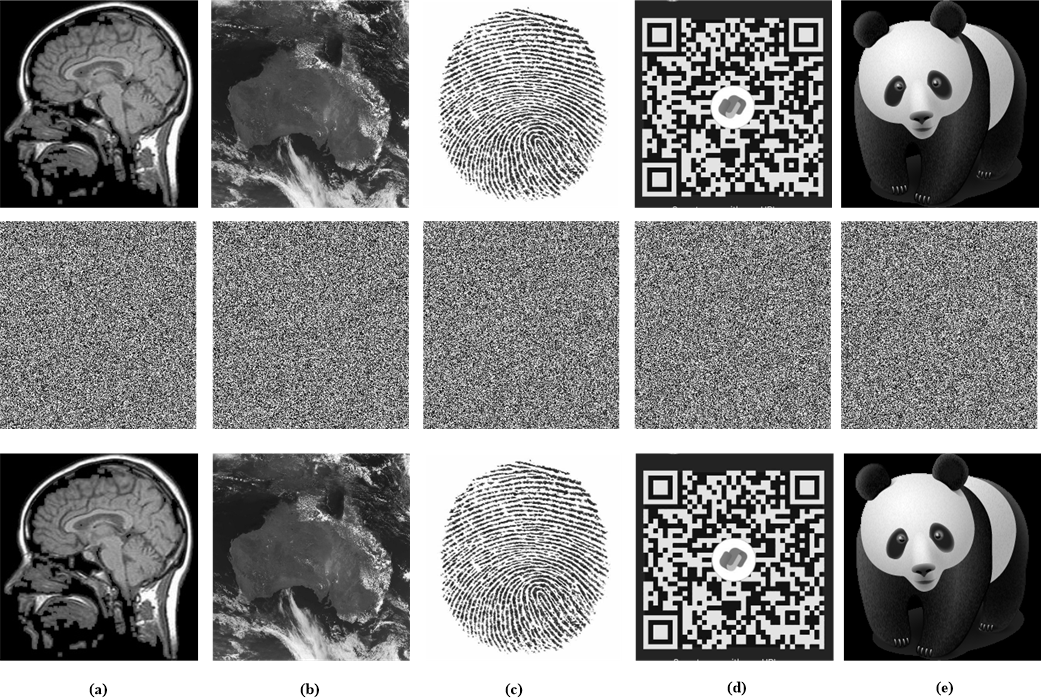}
\caption{Original images (first row) and their corresponding encrypted images (second row) and decrypted images ( third row) of (a) Medical,(b) Satellite,(c) Finger-print,(d) Q-R code,(e) Panda}
\label{Fig-er}
\end{figure}

In the previous sections the encryption and decryption algorithms for different test images were described in detail. The proposed cryptosystem was implemented in 
Python and tested on a laboratory desktop equipped with an Intel i5 processor and 16~GB RAM running the Windows~10 operating system. The experimental results are presented in Fig.~\ref{Fig-er}.

\section{\bf Security and Statistical Analysis}\label{sa}
To evaluate the strength of the proposed encryption scheme, a comprehensive set of statistical and security analyses were performed. These include histogram uniformity, entropy measurement, pixel change rate metrics (NPCR and UACI), correlation coefficients, PSNR, structural similarity index (SSIM), and texture feature analysis using the Gray-Level Co-occurrence Matrix (GLCM). Together, these metrics assess how effectively the encryption conceals the original image information and ensuring high resilience against various attacks.
\subsection{Information Entropy}
Entropy quantifies the level of uncertainty or randomness within a system. In the context of image encryption, it indicates how unpredictable the distribution of pixel values is. For an 8-bit grayscale image, the maximum entropy is 8. An encryption algorithm is considered secure if the entropy of the encrypted image is close to this upper limit, implying a high degree of randomness. The entropy of an image depends on its dimensions, smaller images inherently exhibit lower entropy due to limited pixel diversity.
Thus, entropy comparisons are meaningful only between images of the same size or when normalized. The entropy $e(c)$ is calculated as follows:
\begin{equation}
e(c) = - \sum p(c_i) \log_2 p(c_i)
\end{equation}
where $p(c_i)$ is the probability of occurrence of the pixel value $c_i$. Table~\ref{Table-entropy} presents the entropy values for various original and  encrypted images.

\begin{table}[h]
\centering
\caption{Entropy of the original and the encrypted images}
\begin{tabular}{c c c}
\hline
\textbf{Image Name} & \textbf{Original } & \textbf{Encrypted} \\
\hline
Medical & 6.020 & 7.997 \\
Satellite & 7.094 & 7.997 \\
Finger-print & 5.517 & 7.997 \\
Q-R code & 4.820 & 7.997 \\
Panda & 5.907 & 7.997\\
\hline
\end{tabular} 
\label{Table-entropy}
\end{table}
\subsection{Histogram Analysis}
\begin{figure}[h]
\centering
\includegraphics[width=0.4\linewidth]{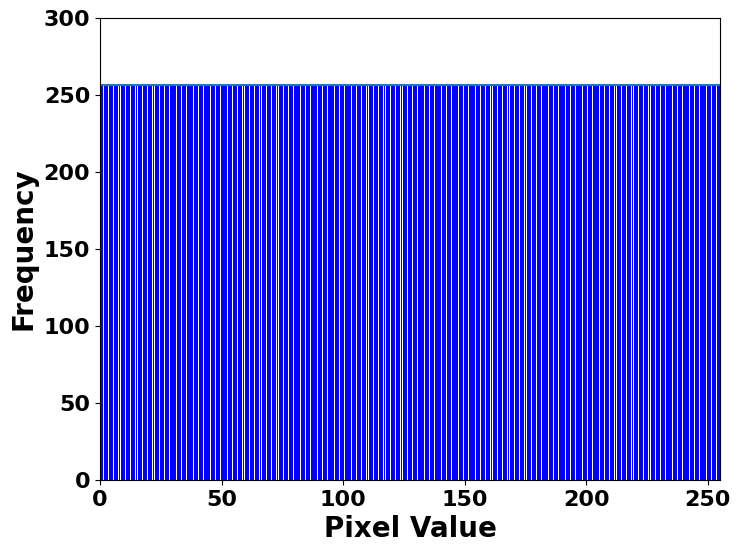}
\caption{Histogram of wheel generated image}
\label{Fig-hwa}
\end{figure}

\begin{figure}[h]
\centering
\includegraphics[width=0.99\linewidth,]{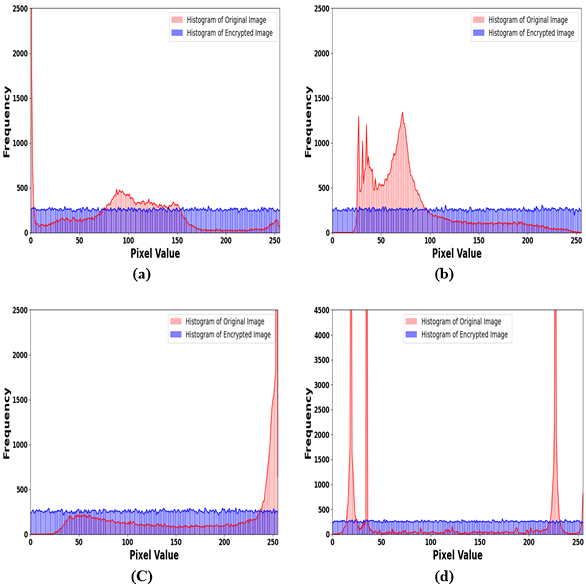}
\caption{Histogram of original (in the red graph) and encrypted (in the blue graph) in (a) Medical,(b) Satellite,(c) Finger-print,(d)Q-R code}
\label{Fig-ha}
\end{figure}

Histograms provide a graphical representation of the statistical distribution of pixel intensities in an image. In an 8-bit grayscale image, pixel values range from 0 to 255. A robust encryption scheme should produce an encrypted image whose histogram appears uniformly distributed, masking all statistical features of the original image. This helps prevent attackers from extracting useful information based on frequency analysis. Figure~\ref{Fig-hwa}, Figure~\ref{Fig-ha} illustrates the histograms for both wheel generated random image, and original and encrypted images respectively. The uniformly distributed histograms of the encrypted images confirm the algorithm’s effectiveness in concealing the original image data.

\subsection{Differential Attack Analysis}

The ability of an encryption scheme to withstand differential analysis is commonly assessed using the Number of Pixel Change Rate (NPCR) and Unified Average Changing Intensity (UACI). These metrics quantify the extent of pixel changes and intensity variation introduced by the encryption process between the original image and its encrypted counterpart.
The NPCR is defined as
\begin{equation}
\text{NPCR} = \frac{1}{M \times N} \sum_{i=1}^{M} \sum_{j=1}^{N} \delta_{ij} \times 100\%
\end{equation}
where
\begin{equation}
\delta_{ij} =
\begin{cases}
1, & \text{if } O_{ij} \ne E_{ij} \\
0, & \text{otherwise}
\end{cases}
\end{equation}
and the UACI is defined as
\begin{equation}
\text{UACI} = \frac{1}{M \times N} \sum_{i=1}^{M} \sum_{j=1}^{N} \frac{|O_{ij} - E_{ij}|}{255} \times 100\%.
\end{equation}

If the encrypted image $E$ is uniformly distributed over $[0,255]$ and statistically independent of the original image $O$, the expected absolute intensity difference at each pixel is
\begin{align}
\mathbb{E}[|O(i,j) - E(i,j)|] 
&= \frac{1}{256} \sum_{k=0}^{255} |O(i,j) - k| \\
&\approx \mathbb{E}[E] \\
&= \frac{0 + 255}{2} = 127.5.
\end{align}

In the special case where $O$ is a pure black image ($O(i,j)=0$), this equality holds exactly:
\begin{equation}
\mathbb{E}[|0 - E(i,j)|] = \mathbb{E}[E] = 127.5.
\end{equation}

Substituting this into the UACI formula gives
\begin{equation}
\mathrm{UACI} \approx \frac{127.5}{255} \times 100\% \approx 50\%.
\end{equation}
Since our test images contain a relatively high proportion of black pixels, the UACI values are higher, approaching the theoretical maximum of 50\%. Similarly for pure white it is 50\%.
\begin{table}[ht]
\centering
\caption{Percentage of Dark or white  Pixels (Intensity $<50$) in Test Images}
\label{tab:dark_pixels}
\begin{tabular}{l c}
\hline
\textbf{Image} & \textbf{Dark/White Pixels (\%)} \\
\hline
Medical & 43.98(D) \\
Satellite & 25.45(D) \\
Fingerprint & 82.40(W) \\
QR-code & 54.30(D) \\
Panda (social media) & 66.93(D) \\
\hline
\end{tabular}
\end{table}

\begin{table}[ht]
\centering
\caption{NPCR and UACI values for different test images}
\begin{tabular}{lcc}
\hline
\textbf{Image} & \textbf{NPCR (\%)} & \textbf{UACI (\%)} \\
\hline
Medical & 99.60 & 36.86 \\
Satellite & 99.60 & 31.75 \\
Finger-print & 99.59 & 41.80 \\
Q-R code & 99.63 & 39.60 \\
Panda (social media image) & 99.60 & 41.60 \\
\hline
\end{tabular}
\label{Table-npcruaci}
\end{table}

\subsection{PSNR}
Peak Signal-to-Noise Ratio (PSNR) widely used metrics to evaluate the visual quality of decrypted images. PSNR is calculated based on the Mean Squared Error (MSE) between the original and reconstructed images:
\begin{equation}
\text{MSE} = \frac{1}{M \times N} \sum_{i=1}^{M} \sum_{j=1}^{N} [O_{ij} - D_{ij}]^2
\end{equation}
\begin{equation}
\text{PSNR} = 10 \log_{10} \left( \frac{O_{\text{max}}^2}{\text{MSE}} \right)
\end{equation}
where $O_{\text{max}}$ is the maximum possible pixel value (255 for 8-bit images), $O$ denotes the original image, and $D$ the decrypted image. The results for encrypted image presented in Table~\ref{tab:psnr_results} confirm the high-quality reconstruction capability of the proposed algorithm.
\begin{table}[H]
\centering
\caption{PSNR Values Between Original, Encrypted, and Decrypted Images}
\label{tab:psnr_results}
\begin{tabular}{c c c}
\hline
\textbf{Image} & \textbf{PSNR (O--E) (dB)} & \textbf{PSNR (O--D) (dB)} \\
\hline
Medical &  6.954 & $\infty$\\
Satellite &  8.208 & $\infty$ \\
Finger-print  & 5.994 & $\infty$ \\
Q-R code &  6.371 & $\infty$ \\
Panda & 6.019 & $\infty$ \\
\hline
\end{tabular}
\end{table}

\subsection{Correlation analysis}
\begin{figure}[h]
\centering
\includegraphics[width=0.90\linewidth]{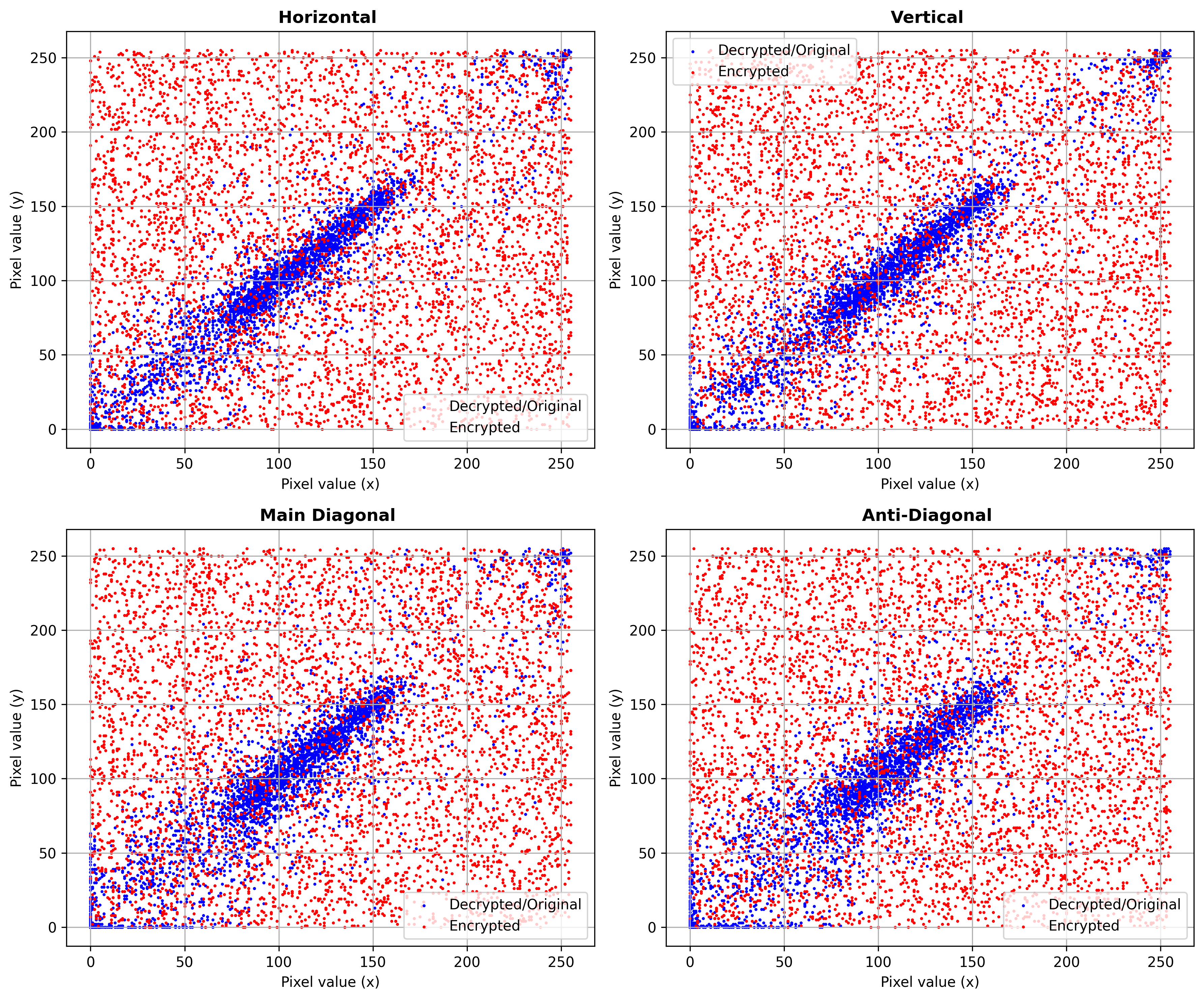}
\caption{Scattered  plot of original (in the blue dots) and encrypted (in the red dots) for medical image in Horizontal, Vertical, Diagonal, Anti Diagonal  directions.}
\label{Fig-sp}
\end{figure}

In original images, adjacent pixels usually exhibit strong correlations 
along horizontal, vertical, and diagonal directions. For an encryption scheme 
to be considered effective, these correlations should be significantly reduced, 
ideally approaching zero, which indicates minimal similarity between neighboring 
pixels in the encrypted image. The correlation coefficient is defined as:  

\begin{equation}
\text{Corr} = \frac{\sum_{i=1}^n (X_i - \bar{X})(Y_i - \bar{Y})}
{\sqrt{\sum_{i=1}^n (X_i - \bar{X})^2} \cdot 
\sqrt{\sum_{i=1}^n (Y_i - \bar{Y})^2}},
\end{equation}

\noindent where $n$ denotes the number of pixel pairs, $X_i$ and $Y_i$ are the gray values of two adjacent pixels, and $\bar{X}$, $\bar{Y}$ are their mean values. The correlation analysis is further illustrated using scatter plots for medical image in Fig.~\ref{Fig-sp} while the numerical values of the correlation coefficients in different directions are reported in Table~\ref{Table-cr}.
\begin{table}[H]
\centering
\caption{Correlation coefficient of original and encrypted images in different directions}
\begin{tabular}{lccccc}
\hline
\textbf{Image} & \textbf{Type} & \textbf{Horizontal} & \textbf{Vertical} & \textbf{Diagonal} & \textbf{Anti-DG} \\
\hline

Medical & Original  & 0.964 & 0.967 & 0.933 & 0.938 \\
       & Encrypted & 0.000 & -0.002 & -0.004 & -0.000 \\

Satellite & Original  & 0.883 & 0.870 & 0.854 & 0.826 \\
          & Encrypted & -0.001 & 0.002 & 0.002 & -0.001 \\

Finger-print & Original  & 0.802 & 0.647 & 0.581 & 0.549 \\
             & Encrypted & 0.001 & -0.003 & 0.000 & -0.001 \\

Q-R code & Original  & 0.897 & 0.891 & 0.799 & 0.801  \\
         & Encrypted & 0.003 & -0.005 & 0.000 & 0.001\\

Panda & Original  & 0.981 &  0.985 & 0.973 &  0.971 \\
      & Encrypted & -0.002 & -0.004 & -0.006 & -0.001 \\

\hline
\end{tabular}
\label{Table-cr}
\end{table}

\subsection{Image Texture Analysis}

Texture provides crucial information about the structural arrangement of surfaces in digital images. One of the most effective statistical approaches for analyzing texture is the Gray-Level Co-occurrence Matrix (GLCM). From this matrix, several statistical texture descriptors can be derived:\\
 \textbf{Dissimilarity:} It measures the variation of gray-level pairs by their absolute difference.
\begin{equation}
    \text{Dissimilarity} = \sum_{i=0}^{N-1} \sum_{j=0}^{N-1} P(i,j)\, | i - j |
 \end{equation}\\
 \textbf{Homogeneity:} It shows the similarity between pixel values, giving higher weight to near-diagonal elements.
    \begin{equation}
        \text{Homogeneity} = \sum_{i=0}^{N-1} \sum_{j=0}^{N-1} \frac{P(i,j)}{1 + |i - j|}
    \end{equation}\\
   \textbf{Contrast:} It quantifies the intensity variation between neighboring pixels.
    \begin{equation}
        \text{Contrast} = \sum_{i=0}^{N-1} \sum_{j=0}^{N-1} (i - j)^2 \, P(i,j)
    \end{equation}\\

  \textbf{Energy:} This measures the textural uniformity.
    \begin{equation}
        \text{Energy} = \sum_{i=0}^{N-1} \sum_{j=0}^{N-1} P(i,j)^2
    \end{equation}

 The outcomes of the texture analysis result are presented in Table~\ref{tab:grayscale_metrics} for clarity and evaluation.\\

\begin{table}[ht]
\centering
\caption{GLCM texture metrics for in original and encrypted medical images}
\label{tab:grayscale_metrics}
\begin{tabular}{l l c c c c}
\hline
\textbf{Image} & \textbf{Type} & \textbf{Dis} & \textbf{Hom} & \textbf{Contrast} & \textbf{Energy} \\
\hline
Medical & Original  & 0.499 & 0.798 & 1.056 & 0.377 \\
     & Encrypted & 5.336 & 0.168 & 42.952 & 0.062 \\
\hline
Satellite & Original  & 0.789 &  0.720 & 2.268 & 0.259 \\
     & Encrypted & 5.317& 0.168 & 42.587 & 0.062 \\
\hline
Finger-print & Original  & 2.186 & 0.601 & 15.160 & 0.484 \\
     & Encrypted & 5.316 & 0.168 & 42.627 & 0.062 \\
\hline
Q-R Code & Original  & 0.912 & 0.853 & 7.760 & 0.430 \\
     & Encrypted & 5.330 & 0.168 & 42.818 & 0.062 \\
\hline
Panda & Original  & 0.230 & 0.915 & 0.726 & 0.412\\
     & Encrypted & 5.349 & 0.166 & 42.994 & 0.062 \\
\hline
\end{tabular}
\end{table}

\subsection{Key Sensitivity}
A secure encryption algorithm must ensure that its output is highly sensitive to the secret key. Even a minimal change, such as altering a single bit or character of a 32-character key, should produce a completely different encrypted key generated by wheel. In our experiments, changing a single character in the key resulted in a Structural Similarity Index Measure (SSIM) of 0.020 between the two encrypted key, indicating almost no structural similarity. This demonstrates that the proposed scheme exhibits strong key sensitivity , ensuring that even tiny key modifications lead to entirely different encryption outputs, which is essential for resisting brute-force and key-related attacks.
\subsection{Brute-Force Resistance}
A brute-force attack systematically tests all possible keys until the correct one is found. The proposed method employs a key space large enough to make this approach computationally infeasible. With a 32-character key chosen from 95 printable ASCII symbols, the total possible combinations are \(95^{32} \approx 1.5 \times 10^{63}\). Even at a rate of \(10^{12}\) key trials per second, an exhaustive search would require approximately \(4.8 \times 10^{43}\) years, vastly exceeding the age of the universe. Even when considering Grover's quantum search algorithm, which reduces the effective complexity to the square root of the key space, the required time would still be on the order of \(10^{32}\) years. This ensures that guessing the key through brute force is practically impossible.
\subsection{Chosen-Plaintext Security}
A chosen plaintext attack (CPA) is a cryptanalytic scenario in which an attacker is allowed to encrypt multiple plaintext images of their choice under the same secret key and then analyze the corresponding ciphertexts to extract information about the encryption mechanism. This attack is particularly effective against image encryption schemes that rely on linear operations or key-stream reuse.
To evaluate the resistance of the proposed encryption scheme against CPA, two different plaintext images were selected and encrypted using the same secret key. Let $P_1$ and $P_2$ denote the plaintext images, and $C_1$ and $C_2$ represent their corresponding encrypted images. If an encryption scheme is vulnerable, the XOR operation between encrypted images may reveal structural similarities corresponding to the XOR of plaintexts, expressed as
\begin{equation}
C_1 \oplus C_2 = P_1 \oplus P_2,
\end{equation}
which indicates key-stream reuse or insufficient diffusion.
In the conducted experiment, XOR operations were performed on two pairs of images: (i) the decrypted versions and (ii) the encrypted versions of two different plaintext images(here finger print and panda taken), and the resulting XOR images were compared pixel-wise. The low SSIM value 0.012 between the two XOR images confirms that the encryption process eliminates linear correlations between plaintexts and ciphertexts, thereby demonstrating strong resistance to chosen plaintext attacks. This nonlinearity arises from a quantum-fluctuation-driven pixel shuffling mechanism, which dynamically alters spatial dependencies and prevents linear propagation of plaintext differences into the ciphertext.
Therefore, the proposed encryption scheme successfully resists chosen plaintext attacks by ensuring strong confusion and diffusion properties. \subsection{Resistance to Intensity Tampering}
\begin{figure}[h]
\centering
\includegraphics[width=.6\linewidth]{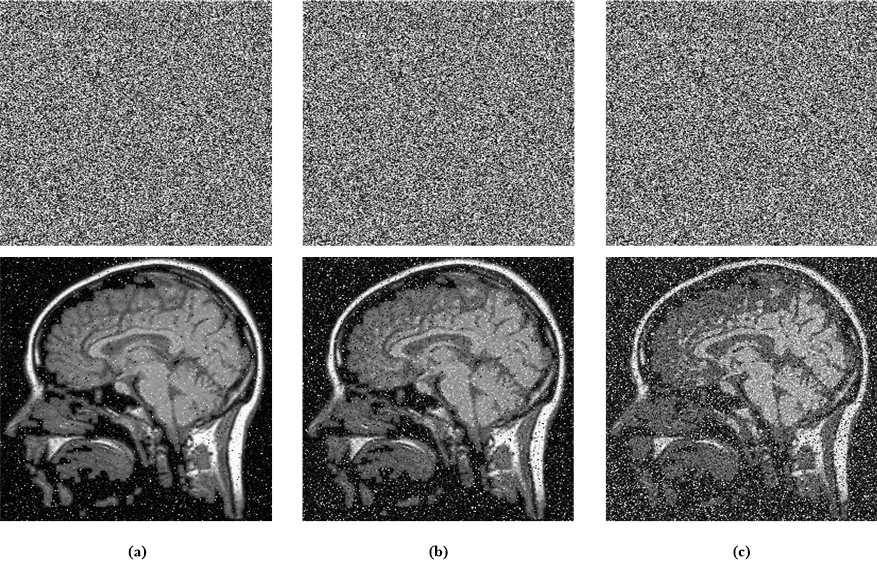}
\caption{{Gaussian noise attack on the encrypted image of  with variance levels: (a) 0.0001, (b) 0.001, (c) 0.01. The first row shows noisy encrypted images, and the second row shows their corresponding decrypted results.}}
\label{Fig-gna}
\end{figure}

\begin{figure}[h]
\centering
\includegraphics[width=.6\linewidth]{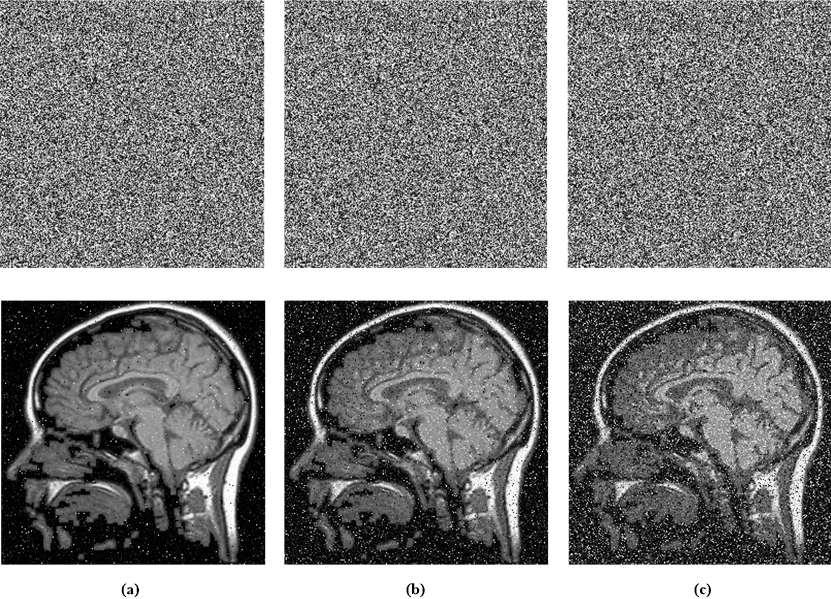}
\caption{{speckel noise attack on the encrypted image of  with variance levels: (a) .01, (b) .05, (c) .15 The first row shows noisy encrypted images, and the second row shows their corresponding decrypted results.}}
\label{Fig-sna}
\end{figure}

The robustness of the proposed method against alterations in pixel intensity was evaluated through noise addition and cropping tests. When the encrypted image is modified by introducing random noise or by removing a portion of its pixels, decryption still produces an image, but with distortions proportional to the degree of alteration. Although perfect reconstruction is not possible in such cases, the ability to recover partially recognizable content demonstrates that the scheme maintains a degree of tolerance to transmission errors and intentional tampering.The effects of noise and cropping attacks are illustrated in Fig.\ref{Fig-gna},and Fig.\ref{Fig-sna}, respectively.

\begin{figure}[h]
\centering
\includegraphics[width=.99\linewidth]{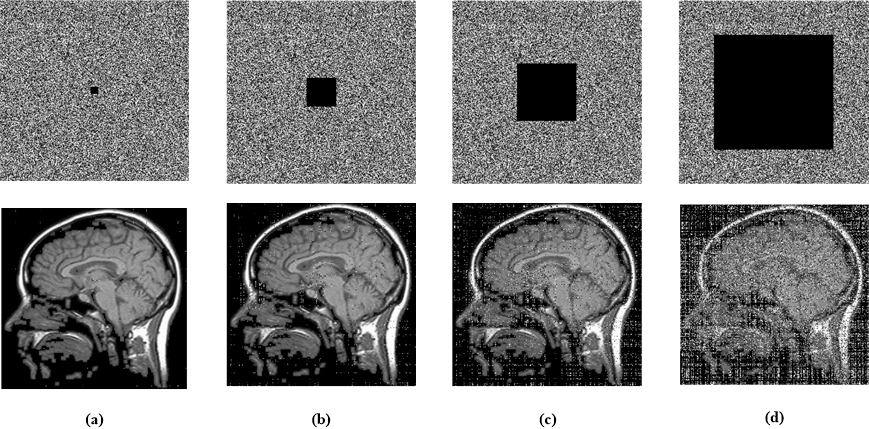}
\caption{{Cropping attack on the encrypted image by removing pixels from the center: (a) $10\times10$, (b) $40\times40$ (c) $80\times80$ (d)$160\times160$. The first row shows cropped encrypted images, and the second row shows the corresponding decrypted images.}}
\label{Fig-ca}
\end{figure}
\subsection{NIST SP 800-22}
The NIST SP 800-22 statistical test suite provides a quantitative framework for evaluating the randomness characteristics of sequences through 15 distinct tests. In this study generated sequences are first transformed into binary streams and then employed as inputs to the test suite. For a sequence to be considered random, the corresponding p-values are required to lie within the interval of 0.01 to 1. As summarized in Table\ref{tab:NIST}, all obtained p values satisfy this criterion, indicating strong randomness properties.
\begin{table*} 
\centering
\caption{NIST SP 800-22 Results}
\label{tab:NIST}
\begin{tabular}{lcc|
cc|cc}
\hline
\multirow{2}{*}{\textbf{Test}} & \multicolumn{2}{c|}{$\mathbf{P_{\text{quantum fluctuation}}}$} & \multicolumn{2}{c|}{$\mathbf{P_{\text{wheel genareted random number}}}$} & \multicolumn{2}{c}{$\mathbf{P_{\text{Encrypted}}}$} \\
 & \textbf{Value} & \textbf{Conclusion} & \textbf{Value} & \textbf{Conclusion} & \textbf{Value} & \textbf{Conclusion} \\
\hline
Frequency & 0.1239 & pass & 1 & pass & 0.1952 & pass \\
Block Freq. & 0.8295 & pass & 0.8525 & pass &0.9460 & pass \\
Runs & 0.2841 & pass &0.8858 & pass &0.2067 & pass \\
Longest Run & 0.8683 & pass & 0.1149 & pass &0.2824 & pass \\
Rank & 0.5716 & pass & 0.6555 & pass &0.6767 & pass \\
DFT & 0.2296 & pass & 0.1492 & pass & 0.3279& pass \\
Non-Overlap Temp. & 0.9335 & pass & 0.7541 & pass & 0.4656 & pass \\
Overlap Temp. & 0.6102 & pass & 0.7502 & pass & 0.5514 & pass \\
Universal & 0.9400 & pass & 0.9619 & pass & 0.9626& pass \\
Appr. Entropy & 0.8402 & pass & 0.9444 & pass & 0.7218& pass \\
Cum. Sums (F) & 0.1292	 & pass & 0.3113 & pass & 0.2207 & pass \\
Cum. Sums (R) & 0.2333 & pass & 0.3113 & pass & 0.3057 & pass \\
Serial & 0.6122 & pass & 0.8519 & pass & 0.4144 & pass \\
Lin. Complexity & 0.4277 & pass & 0.1197 & pass & 0.6875  & pass \\
\hline
\end{tabular}
\end{table*}
\subsection{Structural Similarity Index Measure (SSIM)}
The Structural Similarity Index Measure (SSIM) is a perceptual metric commonly employed to evaluate the visual fidelity of decrypted images. Unlike conventional error-based measures such as Mean Squared Error (MSE) or Peak Signal-to-Noise Ratio (PSNR), which rely solely on pixel-wise differences, SSIM quantifies similarity by jointly considering luminance, contrast, and structural information. For two images $x$ and $y$, SSIM is expressed as:

\[
\text{SSIM}(x, y) = \frac{(2\mu_x \mu_y + C_1)(2\sigma_{xy} + C_2)}{(\mu_x^2 + \mu_y^2 + C_1)(\sigma_x^2 + \sigma_y^2 + C_2)}
\]

where $\mu_x$ and $\mu_y$ denote the mean intensity values, $\sigma_x^2$ and $\sigma_y^2$ represent the variances, $\sigma_{xy}$ is the covariance between the images, and $C_1$ and $C_2$ are small constants introduced to ensure numerical stability. The SSIM value lies in the range $[0,1]$, where a value of 1 indicates perfect similarity. In the context of image encryption, a high SSIM value between the original and decrypted images confirms accurate reconstruction and minimal perceptual distortion. Owing to its strong correlation with human visual perception, SSIM is widely adopted for evaluating both image quality and cryptographic performance. The corresponding results are reported in Table~\ref{tab:ssim}.

\begin{table}[ht]
\centering
\caption{SSIM between original–encrypted (O–E) and original–decrypted (O–D) images}
\label{tab:ssim}
\begin{tabular}{lcc}
\hline
\textbf{Image} & \textbf{O--E} & \textbf{O--D} \\
\hline
Medical        & 0.007 & 1.000 \\
Satellite      & 0.009 & 1.000 \\
Finger-print   & 0.004 & 1.000 \\
Q-R code       & 0.002 & 1.000 \\
Panda          & 0.005 & 1.000 \\
\hline
\end{tabular}
\end{table}
\section{Security parameter Analysis Under Intensity Tampering Attack }
 A comprehensive evaluation of the proposed encryption scheme under intensity tampering attacks are  presented in Table~\ref{tab:crop_parameter}. In case of cropping attack we have consider cropping size of 10 × 10, 40 × 40, 80 × 80, and 160 × 160 as denoted by $CA-1$ to $CA-4$ respectively. We have also include noise attacks, namely Gaussian  and spackel noise. Gaussian noise correspond to $GNA-1$ to $GNA-3$    with variance levels
 0.0001, 0.001, 0.01 and speckel noise attacks $SNA-1$ to $SNA-3$  with variance levels
 0.01, 0.05, 0.15 respectively. These attacks simulate realistic transmission distortions and tampering scenarios commonly encountered in image communication.\\ The PSNR(O–E) values remain low (6–7 dB) across all attack scenarios, indicating significant visual dissimilarity between the original and encrypted images, which is essential for confidentiality. Meanwhile, PSNR(O–D) values decrease with increasing attack severity, particularly under aggressive cropping and high-density noise, reflecting controlled degradation during decryption rather than catastrophic failure. This behavior is desirable for consumer electronics applications, where partial image recovery under tampering is preferable to complete information leakage. Moreover, the SSIM values degrade gradually with attack severity. Overall, the experimental results demonstrate that the proposed encryption scheme maintains strong security and robustness under various intensity tampering attacks, making it highly suitable for secure image transmission in adversarial and noisy environments.

\begin{table*}

\centering
\caption{Security parameter for medical image Analysis Under Cropping and Noise Attacks}
\label{tab:crop_parameter}
\begin{tabular}{c c c c c c c c c c c c}
\hline
Parameter & Original & CA-1 & CA-2 & CA-3 & CA-4 & GNA-1 & GNA-2 & GNA-3 & SNA-1 & SNA-2 & SNA-3 \\
\hline
NPCR & 99.60 &99.59  & 99.43  & 99.15 & 98.55 & 99.57 & 99.62 &99.63  & 99.62 & 99.63  & 99.59  \\
UACI & 36.86 & 36.87  & 37.08  & 37.82  & 41.16  & 35.77 & 35.08  & 34.37  & 36.17 & 35.31  & 34.53 \\
Horizontal(CC)& 0.000  & 0.004 & 0.066  & 0.221  & 0.526  & 0.000  & 0.000  & 0.000  & 0.000 & 0.000 & 0.000   \\
Vertical(CC) &-0.002 & -0.002 & 0.060 & 0.217 &0.526  &-0.002  &-0.003 &-0.003  & -0.003 & -0.005  & -0.005    \\
Diagonal(CC) &-0.004  & -0.001 & 0.059  & 0.213  &0.519  & -0.004 &0.005  & 0.005 & -0.004  &-0.004  &-0.004     \\
Anti Diagonal(CC) &-0.000 & 0.002 & 0.064  & 0.218  &0.524  & 0.000 &-0.000 &-0.000  &-0.000  &-0.001  &-0.002  \\
PSNR(O-E) & 6.954 & 6.954  & 6.922 & 6.820 & 6.279  & 6.973  &6.964  & 6.875 & 7.099  & 7.293  & 7.454 \\
PSNR(O-D) & $\infty$ & 34.74 & 23.04 & 17.18 & 11.07  & 20.16 & 16.34  &12.30  & 23.47 & 17.35  & 13.20 \\
Dissimilarity & 0.499 & 0.516  & 0.735  & 1.353 & 3.384  & 1.066  & 1.793  & 5.486 & 0.760  & 1.552 & 2.794  \\
Homogeneity & 0.798  & 0.795 & 0.768  & 0.690 & 0.435  &0.678  & 0.532  &0.163  &0.742 & 0.581  & 0.387 \\
Contrast & 1.056 & 1.213 & 3.215  & 8.713 & 26.861 & 4.899 & 9.958 & 45.411  & 2.872  & 8.343 & 18.212 \\
Energy &  0.377 & 0.376 & 0.361  &  0.321 & 0.185 &0.283 & 0.187  &0.060  &0.333 & 0.217 & 0.121  \\
SSIM & 1.000 & 0.979  & 0.769  & 0.458 & 0.167 & 0.524 & 0.384 &0.001  & 0 .633 & 0.424 & 0.258  \\
\hline
\end{tabular}
 \end{table*}
\subsection{Ablation Study}
To analyze the contribution of each major component of the proposed encryption framework, an ablation study was conducted. In the ablation study we selectively disabling one module at a time while keeping the remaining stages unchanged. The impact of each stages in the overall encryption algorithm are discussed below.
\subsubsection{A1: Removal of image Shuffling}
\begin{figure}[H]
\centering
\includegraphics[width=.4\linewidth]{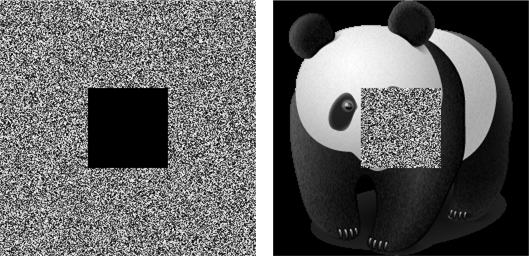}
\caption{{Cropping attack without image shuffling.}}
\label{Fig-Aca}
\end{figure}

In this ablation part, the initial image shuffling stage is removed, and the plaintext image is directly passed to the diffusion and substitution processes. As shown in Fig.~\ref{Fig-Aca}, the absence of shuffling weakens the encryption algorithm. In this case the cropping attacks  partially successful. Similarly chosen-plaintext attacks become more easy, as linear relationships between original and encrypted image pixels exist. It is easier for an attacker to predict diffusion patterns. These observations confirm that pixel shuffling is critical for breaking spatial redundancy and providing strong confusion, thereby enhancing the overall security of the proposed scheme.
\subsubsection{A2: Without Quantum Fluctuation--Based Sequence}
In A2, the quantum kicked rotor generated energy sequence is replaced with a seed any arbitrary number (215) generated random sequence. This modification significantly reduces entropy and increased corelation resulted in Table-\ref{tab:correlation_entropy}, indicating weakened randomness and diffusion. The results confirm that quantum fluctuation driven dynamics play a crucial role in enhancing unpredictability and strengthening resistance to differential attacks.
\begin{table}
\centering
\caption{Correlation Coefficients (Horizontal, Vertical, Diagonal, Anti-diagonal) and Entropy Comparison of Original and Ablation Variants}
\label{tab:correlation_entropy}
\begin{tabular}{c c c c c c}
\hline
\textbf{Variant} & \textbf{H} & \textbf{V} & \textbf{D} & \textbf{A-D} & \textbf{Ent} \\
\hline
Panda(Propose method) &-0.002 & -0.004 & -0.006 & -0.001 & 7.997\\
A2& 0.004 & 0.006 & -0.011 &-0.007 & 7.984 \\
\hline
\end{tabular}
\end{table}

\subsubsection{A3: Static Rotating Wheel}
In this ablation variant, the dynamic shuffled wheel is disabled, resulting in a static wheel during encryption. Consequently, the main encryption key, which is derived from wheel, exhibits significantly increased correlation between successive key stream elements. This indicates reduced randomness and leads to periodic in nature. The results demonstrated in Table-\ref{tab:correlation_abalation} that the dynamic shuffled wheel is essential for breaking correlations, avoiding predictable patterns, and enhancing temporal randomness, thereby maintaining the overall security and unpredictability of the proposed encryption scheme.
\subsubsection{A4: Single-Tap Sampling}
In A4, the multi-tap sampling strategy is simplified to a single tap per iteration. The statistical results in Table \ref{tab:correlation_abalation} indicate a tendency toward higher correlation among generated values and a corresponding decline in performance.
\begin{table}[H]
\centering
\scriptsize
\caption{Directional Correlation Comparison of Original and Ablation Variants. H: Horizontal, V: Vertical, D: Diagonal, A-D: Anti-diagonal}
\label{tab:correlation_abalation}
\begin{tabular}{l c c c c}
\hline
\textbf{Variant} & \textbf{H} & \textbf{V} & \textbf{D} & \textbf{A-D} \\
\hline
Wheel generated random number (P.M) & -0.027 & -0.018 & -0.022 & -0.013 \\
A3: Without shuffled wheel & -0.185 & -0.031 & 0.036 & -0.024 \\
A4: Single tap & -0.027 & -0.097 & -0.044 & -0.014 \\
\hline
\end{tabular}
\end{table}
Overall, the ablation results confirm that each module contributes meaningfully to the security and robustness of the proposed encryption scheme, and optimal cryptographic performance is achieved only when all components operate jointly.
\begin{table}[H]
\scriptsize
\setlength{\tabcolsep}{2pt}
\centering
\caption{Performance Comparison of Image Encryption Techniques}
\begin{tabular}{lcccccccc}
\hline
Metric & Proposed(CPU) & \cite{bib120} & \cite{bib123} & \cite{bib122} & \cite{bib121} & AES \\
\hline
Entropy         & 7.997 & 7.997 & 7.997 & 7.998 & 7.990 & 7.869\\
NPCR (\%)       & 99.60 & 99.22 & 99.58 & 99.58 & 99.60 & 99.40 \\
UACI (\%)   & 36.86 & 33.10 & 33.54 & 33.37 & 33.41 & 28.00 \\
CC (Horizontal) & 0.000 & 0.003 & 0.003 & 0.001 & 0.003 & 0.270\\
CC (Vertical)   & -0.002 & 0.007 & 0.007 & - & -0.003 & 0.268 \\
CC (Diagonal)   & -0.004 & 0.004 & 0.001 & - & 0.006 & 0.075 \\
\hline
\end{tabular}
\label{tab:comparison}
\end{table}
\section{Comparison with related work}
Performing a direct comparison of different encryption studies is difficult due to variations in image characteristics such as content, size, and texture. However, approximate evaluations can still reveal performance trends across encryption methods. Such comparisons must be interpreted with caution, as differences in image properties can heavily impact the outcomes. Table~\ref{tab:comparison} compares key metrics of the proposed encryption method against other techniques.
\section{Conclusion}
This paper presents a new random number generator based on quantum inspired fluctuations and a rotating wheel mechanism. The generator uses only integer operations, which makes it stable and independent of floating-point precision. Quantum-inspired data are first used to shuffle the wheel. Then, dynamic wheel rotation and key-dependent multi-tap sampling are applied. Because of this design, the generated random numbers show ideal entropy and low correlation. The proposed random number generator is then applied to image encryption. The generated sequences are used for pixel permutation and diffusion, which ensures strong confusion and diffusion. Extensive experimental results, including entropy, NPCR, UACI, correlation, PSNR, and SSIM analyses, confirm strong security performance. Robustness against noise, cropping, brute-force, and chosen-plaintext attacks is also verified.  Overall, the proposed method offers a simple, efficient, and secure image encryption solution. The proposed random number generator relies on tap positions and tap count, for which no universally optimal configuration exists; consequently, inappropriate selections can adversely affect statistical performance is the main limitation. Future work will focus on optimizing the algorithm for faster execution and lower memory usage, and on applying the proposed random number generator to other security applications such as video encryption and secure data transmission.


\begin{thebibliography}{1}
\bibliographystyle{IEEEtran}
\bibitem{bib1}W.~Alexan, N.~H.~El~Shabasy, N.~Ehab, and E.~A.~Maher,
``A secure and efficient image encryption scheme based on chaotic systems and nonlinear transformations,''
\emph{Sci. Rep.}, vol.~15, no.~1, Art.~no.~31246, 2025.
\bibitem{bib2}
P.~L.~Sharma, S.~Gupta, H.~Monga, A.~Nayyar, K.~Gupta, and A.~K.~Sharma,
``TEXCEL: Text encryption with elliptic curve cryptography for enhanced security,''
\emph{Multimedia Tools Appl.}, vol.~84, no.~13, pp.~11503--11531, 2025.
\bibitem{bib3}
J.~Daemen and V.~Rijmen,
``AES proposal: Rijndael,''
Natl. Inst. Stand. Technol. (NIST), Gaithersburg, MD, USA, 1999.
\bibitem{bib4}
D.~Coppersmith,
``The data encryption standard (DES) and its strength against attacks,''
\emph{IBM J. Res. Develop.}, vol.~38, no.~3, pp.~243--250, 1994.
\bibitem{bib5}
N.~Iqbal, A.~Banga, N.~Innab, B.~M.~El~Zaghmouri, A.~Ikram, and H.~Diab,
``Utilizing the $n$th root of numbers for novel random data calculus and its applications in network security and image encryption,''
\emph{Expert Syst. Appl.}, vol.~265, Art.~no.~125992, 2025.
\bibitem{bib6}
M.~Liu, R.~Shaydulin, P.~Niroula, M.~DeCross, S.-H.~Hung, W.~Y.~Kon, E.~Cervero-Mart\'{\i}n, K.~Chakraborty, O.~Amer, S.~Aaronson, \emph{et~al.},
``Certified randomness using a trapped-ion quantum processor,''
\emph{Nature}, pp.~1--6, 2025.
\bibitem{bib7}
A.~Hadj~Brahim, H.~Ali~Pacha, M.~Naim, and A.~Ali~Pacha,
``A novel pseudo-random number generator: Combining hyperchaotic system and DES algorithm for secure applications,''
\emph{J. Supercomput.}, vol.~81, no.~1, Art.~no.~94, 2025.
\bibitem{bib8}
J.~Wu, A.~Y.~Salim, E.~Elmitwalli, S.~K{\"o}se, and Z.~Ignjatovic,
``A pseudo-random number generator for multi-sequence generation with programmable statistics,''
in \emph{Proc. IEEE Int. Symp. Circuits Syst. (ISCAS)}, pp.~1--4, 2025.
\bibitem{bib9}
M.~Stip{\v{c}}evi{\'c} and {\c{C}}.~K.~Ko{\c{c}},
``True random number generators,''
in \emph{Open Problems in Mathematics and Computational Science}, pp.~275--315. Springer, 2014.

\bibitem{bib10}
F.~Yu, L.~Li, Q.~Tang, S.~Cai, Y.~Song, and Q.~Xu,
``A survey on true random number generators based on chaos,''
\emph{Discrete Dyn. Nat. Soc.}, vol.~2019, Art.~no.~2545123, 2019.
\bibitem{bib11}
S.~Rana, A.~Pathak, H.~Mondal, and M.~K.~Mandal,
``Secure healthcare data management with lossless compression and hyperchaos encryption,''
in \emph{Proc. IEEE Calcutta Conf. (CALCON)}, pp.~1--4, 2024.
\bibitem{bib12}
S.~Rana, A.~Pathak, H.~Mondal, and M.~K.~Mandal,
``Region-based medical image encryption for patient privacy,''
\emph{Int. J. Signal Imaging Syst. Eng.}, 2025.
\bibitem{bib13}
B.~Acharya, J.~V.~Sravan, D.~J.~R.~Potnuru, and K.~A.~K.~Patro,
``MIE-SPD: A new and highly efficient chaos-based multiple image encryption technique with synchronous permutation diffusion,''
\emph{IEEE Access}, 2025.
\bibitem{bib14}
H.~Noura, L.~Sleem, and R.~Couturier,
``A revision of a new chaos-based image encryption system: Weaknesses and limitations,''
\emph{arXiv preprint arXiv:1701.08371 [cs.CR]}, 2017. [Online]. Available: \url{https://doi.org/10.48550/arXiv.1701.08371}
\bibitem{bib15}
X.~Nie and W.~Zheng,
``Fate of the spatial-temporal order under quantum fluctuation,''
\emph{Phys. Rev. A}, vol.~111, no.~6, Art.~no.~063313, 2025.
\bibitem{bib135}
G. Casati, B. V. Chirikov, and I. Guarneri,
``Casati, Chirikov, and Guarneri Respond,''
\textit{Phys. Rev. Lett.}, vol. 56, no. 25, p. 2768, 1986.
\bibitem{bib134}
M. S. Santhanam, S. Paul, and J. Bharathi Kannan,
``Quantum kicked rotor and its variants: Chaos, localization and beyond,''
\textit{Physics Reports}, vol. 956, pp. 1--87, 2022.

\bibitem{bib123}
T. S. Monteiro, P. A. Dando, N. A. C. Hutchings, and M. R. Isherwood,
``Proposal for a chaotic ratchet using cold atoms in optical lattices,''
\textit{Physical Review Letters}, vol. 89, no. 19, p. 194102, 2002.
\bibitem{bib124}
P. H. Jones, M. M. Stocklin, G. Hur, and T. S. Monteiro,
``Atoms in double-$\delta$-kicked periodic potentials: chaos with long-range correlations,''
\textit{Physical Review Letters}, vol. 93, no. 22, p. 223002, 2004.
\bibitem{bib125}
S. Paul, J. Bharathi Kannan, and M. S. Santhanam,
``Interaction-induced directed transport in quantum chaotic subsystems,''
\textit{Physical Review E}, vol. 108, no. 4, p. 044208, 2023.
\bibitem{bib126}
S. Paul, H. Pal, and M. S. Santhanam,
``Barrier-induced chaos in a kicked rotor: Classical subdiffusion and quantum localization,''
\textit{Physical Review E}, vol. 93, no. 6, p. 060203, 2016.
\bibitem{bib127}
S. Paul and M. S. Santhanam,
``Floquet states of a kicked particle in a singular potential: Exponential and power-law profiles,''
\textit{Physical Review E}, vol. 97, no. 3, p. 032217, 2018.
\bibitem{bib128}
J. Gong and P. Brumer,
``Coherent control of quantum chaotic diffusion,''
\textit{Physical Review Letters}, vol. 86, no. 9, p. 1741, 2001.
\bibitem{bib129}
J. Gong, H. J. W{\"o}rner, and P. Brumer,
``Control of dynamical localization,''
\textit{Physical Review E}, vol. 68, no. 5, p. 056202, 2003.
\bibitem{bib130}
J. Gong, H. J. W{\"o}rner, and P. Brumer,
``Control of dynamical localization,''
\textit{Physical Review E}, vol. 68, no. 5, p. 056202, 2003.
\bibitem{bib131}
B. G. Klappauf, W. H. Oskay, D. A. Steck, and M. G. Raizen,
``Observation of noise and dissipation effects on dynamical localization,''
\textit{Physical Review Letters}, vol. 81, no. 6, p. 1203, 1998.
\bibitem{bib132}
S. Sarkar, S. Paul, C. Vishwakarma, S. Kumar, G. Verma, M. Sainath, U. D. Rapol, and M. S. Santhanam,
``Nonexponential decoherence and subdiffusion in atom-optics kicked rotor,''
\textit{Physical Review Letters}, vol. 118, no. 17, p. 174101, 2017.
\bibitem{bib133}
S. Sarkar, M. S. Santhanam, J. Mangaonkar, U. D. Rapol, C. Vishwakarma, and S. Paul,
``Nonmonotonic diffusion rates in an atom-optics L\'evy kicked rotor,''
\textit{Physical Review E}, 2019.

\bibitem{bib120}
W. Feng, K. Zhang, J. Zhang, X. Zhao, Y. Chen, B. Cai, Z. Zhu, H. Wen, and C. Ye, 
``Integrating fractional-order Hopfield neural network with differentiated encryption: achieving high-performance privacy protection for medical images,'' 
\textit{Fractal and Fractional}, 
vol. 9, no. 7, p. 426, 2025.

\bibitem{bib121}
C. Lakshmi, K. Thenmozhi, J. B. B. Rayappan, S. Rajagopalan, R. Amirtharajan, and N. Chidambaram, 
``Neural-assisted image-dependent encryption scheme for medical image cloud storage,'' 
\textit{Neural Computing and Applications}, vol. 33, pp. 6671--6684, 2021.
\bibitem{bib123}
P. Sarosh, S. A. Parah, and G. M. Bhat, 
``An efficient image encryption scheme for healthcare applications,'' 
\textit{Multimedia Tools and Applications}, vol. 81, no. 5, pp. 7253--7270, 2022.

\bibitem{bib122}
J.-Y. Dai and N.-R. Zhou,
``Optimal quantum image encryption algorithm with the QPSO-BP neural network-based pseudo random number generator,''
\textit{Quantum Information Processing}, vol. 22, art. no. 318, 2023.


\end{thebibliography}
\end{document}